\documentclass[conference]{IEEEtran}
\ifCLASSINFOpdf
\else
\fi
\usepackage{amsmath}
\usepackage{amssymb}
\usepackage{mathtools}
\usepackage{algorithmic}
\usepackage[mathscr]{euscript}
\usepackage{graphicx,subfigure}
\usepackage{amsthm}
\usepackage[scaled]{helvet} 
\usepackage{cases}
\usepackage{cite}
\usepackage{algorithmic}
\usepackage{epstopdf}
\usepackage{color}
\usepackage[lined,boxed,commentsnumbered,ruled,vlined]{algorithm2e}
\usepackage{makeidx}
\usepackage[]{algorithm2e}
\usepackage{arydshln,leftidx}
\usepackage{blkarray}
\usepackage{xcolor}

\newcommand\bovermat[2]{%
  \makebox[0pt][l]{$\smash{\overbrace{\phantom{%
    \begin{matrix}#2\end{matrix}}}^{\text{#1}}}$}#2}

\newcommand{\Nofdm}{N_o}

\newcommand{\nonr}{\nonumber}
  
\newcommand{\beq}{\begin{equation}} 
\newcommand{\eeq}{\end{equation}} 
\newcommand{\beqa}{\begin{eqnarray}}
\newcommand{\eeqa}{\end{eqnarray}}

\newcommand{\smpMat}{\mathbf{A}}

\newcommand{\mb}{\mathbf{b}}
\newcommand{\atilde}{\tilde{\mb}}

\newcommand{\sS}{\mathscr{S}}
\newcommand{\mBX}{\mathbf{B}} 
\newcommand{\mBdy}{{\mBX}_{d,1}} 
\newcommand{\mBdd}{{\mBX}_{d,2}} 
\newcommand{\mBdys}{{\mBX}_{d,1,\sS}} 
\newcommand{\mBdds}{{\mBX}_{d,2,\sS}} 
\newcommand{\mBds}{{\mBX}_{d,\sS}}

\newcommand{\mP}{\mathbf{P}}

\newcommand{\cC}{\mathcal{C}}

\newcommand{\vs}{\mathbf{s}}

\newcommand{\vz}{\mathbf{z}}
\newcommand{\vw}{\mathbf{w}}
 
\newcommand{\mF}{\mathbf{F}}

\newcommand{\mI}{\mathbf{I}}

\newcommand{\mZero}{\mathbf{0}}

\newcommand{\Ncp}{N_{\mbox{\tiny CP}}}
\newcommand{\Nb}{N_b}

\newcommand{\rank}{\mbox{rank}}

\newcommand{\Ftilde}{\tilde{\mF}} 
\newtheorem{thm}{Theorem}

\newtheorem{lem}{Lemma}

\newcommand{\mB}{\mathbf{B}}
\newcommand{\mA}{\mathbf{A}}

\newcommand{\vx}{\mathbf{x}}

\newcommand{\spark}{\mathrm{spark}}

\newcommand{\myvhskip}{\vskip-6pt \hskip+10pt}

\newcommand{\mLambda}{\mbox{\boldmath$\Lambda$\unboldmath}}

\newcommand{\mPhi}{\mbox{\boldmath$\Phi$\unboldmath}}
\newcommand{\mPsi}{\mbox{\boldmath$\Psi$\unboldmath}}

\newcommand{\kK}{\mathcal{K}}

\makeatletter
\renewcommand*\env@matrix[1][*\c@MaxMatrixCols c]{%
  \hskip -\arraycolsep
  \let\@ifnextchar\new@ifnextchar
  \array{#1}}
\makeatother

\begin{document}
%
\title{Compressive Identification of Active OFDM Subcarriers in Presence of Timing Offset}

\author{\IEEEauthorblockN{Alireza Razavi$^\dagger$, Mikko Valkama$^\dagger$, Danijela Cabric$^\ddagger$}
\IEEEauthorblockA{$^\dagger$Department of Electronics and Communications Engineering, Tampere University of Technology, Tampere, Finland\\
$^\ddagger$CORES Lab., University of California, Los Angeles (UCLA), USA \\
email: alireza.razavi@tut.fi, mikko.e.valkama@tut.fi, danijela@ee.ucla.edu
}}
\maketitle

\begin{abstract}
In this paper we study the problem of identifying active subcarriers in an OFDM signal from compressive measurements sampled at sub-Nyquist rate. The problem is of importance in Cognitive Radio systems when secondary users (SUs) are looking for available spectrum opportunities to communicate over them while sensing at Nyquist rate sampling can be costly or even impractical in case of very wide bandwidth. We first study the effect of timing offset and derive the necessary and sufficient conditions for signal recovery in the oracle-assisted case when the true active sub-carriers are assumed known. Then we propose an Orthogonal Matching Pursuit (OMP)-based joint sparse recovery method for identifying active subcarriers when the timing offset is known. Finally we extend the problem to the case of unknown timing offset and develop a joint dictionary learning and sparse approximation algorithm, where in the dictionary learning phase the timing offset is estimated and in the sparse approximation phase active subcarriers are identified. The obtained results demonstrate that active subcarrier identification can be carried out reliably, by using the developed framework.
\end{abstract}


\section{Introduction}
\label{sec:intro}
Cognitive radio systems are emerging to respond to the ever-increasing demand for higher data rates and new wireless services by solving the spectrum underutilization problem \cite{cr99, haykin05}.
The most vital task in CR systems is Spectrum Sensing (SS) defined as identifying spectrum holes by sensing the radio spectrum and utilizing them 
without causing interference to primary users (PUs) \cite{haykin09}. 
Of special interest in this regard is sensing of OFDM signals \cite{sachin09, axell11}. OFDM is one of the most effective multicarrier techniques for broadband wireless communications which is employed by
many of the current and emerging wireless technologies.

\myvhskip On the other hand, due to the limitations of today's analog-to-digital converter (ADC) circuits which cannot support very high bandwidth and need excessive memory and prohibitive
energy costs for implementing digital signal processing systems \cite{cohen11}, it may be very costly and even impractical to sense the signal based on Nyquist-rate
samples. 
This has motivated researchers to study sub-Nyquist methods for wideband spectrum sensing in CR networks; see, e.g., \cite{mishali11, tian07, tian12, polo09, leus11, ariananda11, cohen11, rebeiz12, razaviSP14}. However the number of works  focusing on OFDM detection from sub-Nyquist samples is very limited. In \cite{razavi13, razavi14} the authors proposed methods for detection of OFDM signals, but because both of these methods need a specific embedded signature inside the signal for the detection algorithm to work, they are suitable for detecting a cognitive network \cite{sutton08} rather than detecting PUs who are not willing to assist in signal detection process by embedding signatures in their signal. Besides, these methods as well as the one proposed in \cite{razaviSP14} do not address the problem of active subcarrier identification and rather focus on the detection of the OFDM signal as a whole. The works in \cite{polo09, ariananda11} reconstruct the power spectral density (PSD) of a signal from its sub-Nyquist measurements which can be used for detecting used subcarriers of OFDM signal. The problem with these methods is that they are designed under the assumption of perfect timing information and 
therefore their performance degrades in the presence of unknown timing offset. Furthermore they need relatively many samples to compute the PSD of sub-Nyquist signal and then reconstruct the one of the original signal.

{\bf Our contribution:} In this paper, we propose a method for identification of active subcarriers of an OFDM signal with unknown timing offset. We assume that the OFDM signal is sampled by an Analog-to-Information converter (AIC) \cite{tropp10, kirolos06} at sub-Nyquist rates. We then model the system as a multiplication of a redundant dictionary which depends on the timing offset and a sparse vector whose non-empty entries correspond to the active subcarriers of OFDM signal. The effect of timing offset is studied and necessary and sufficient conditions for signal recovery in the oracle assisted case when the support of sparse vector is assumed known are provided. This is of great importance as each iteration of the proposed greedy algorithm involves a signal recovery step for the vector support estimated in that iteration. We then propose an algorithm that alternates between two steps: a dictionary learning step during which the timing offset is estimated and a joint sparse recovery step during which the active subcarriers are estimated. The OMP algorithm \cite{omp93,tropp07} is tailored to recover the sparsity pattern of OFDM symbols.

{\bf Paper organization:} The rest of the paper is organized as follows. In Section \ref{sec:2} the system model is given and the problem is formulated. Sections \ref{sec:delay} and \ref{sec:3} are the main parts of the paper. In Section \ref{sec:delay}, we first study the oracle-assisted case, then discuss the issues arising from very small or very large timing offset and introduce a solution for these cases. Then in Section \ref{sec:3} we propose a joint compressive sensing method for identification of active subcarriers when the timing offset is known. We also extend the method to the case of unknown timing offset by employing a joint dictionary learning and sparse approximation algorithm to identify the active subcarriers. The performance of the proposed method will be studied by simulation experiments in Section \ref{sec:simulation}. Conclusions are drawn in Section \ref{sec:conclusion}. 

{\bf Notations and Mathematical Preliminaries:} Throughout this paper matrices and vectors are denoted by capital and small boldface letters, respectively. $=$ denotes the equality and $\triangleq$ denotes the definition.
$\mI$ and $\mZero$ represent, respectively,
identity matrix and all-zero matrix. For a matrix $\smpMat$, $[\smpMat]_{i,j}$ denotes the $(i,j)$-th entry of the matrix and $\mA_\sS$ denotes the sub-matrix of $\mA$ composed of only columns determined by index set $\sS$. The rank, column space, and null space (kernel) of $\mA$ are denoted by $\rank(\mA)$, $\cC(\mA)$, and $\kK(\mA)$, respectively.The set of integers from $p$ to $q>p$ (including $p$ and $q$) is denoted as $\{p:q\}$. For a vector $\vx$, the (sub-)vector consisting of entries determined by index set $\sS$ is denoted as $\vx_\sS$. We denote the empty set by $\emptyset$ and for a set $\sS$ we denote its cardinality by $|\sS|$. 
 
Consider an $M \times N$ matrix $\mA$, the {\it spark} of $\mA$, denoted by $\spark(\mA)$, is the smallest number $n$ such that there exist a sub-set of $n$ columns of $\mA$ which are linearly dependent \cite{donoho03,eladBook}. The following inequality describes the relaltionship between rank and spark \cite{donoho03}:
 \beq
 \spark(\mA) \le \rank(\mA) + 1.
 \label{ineq:rankspark}
 \eeq
If $M \ge N$ and $\mA$ is full-column rank (and therefore there is no subset of columns that are linearly dependent) then we use the convention $\spark(\mA)=N+1$. On the other hand, if $M \le N$ and every $M \times M$ subset of $\mA$ is invertible, then $\mA$ is called a full-spark matrix \cite{fullspark}, i.e. $\spark(\mA)=M+1$. 

\section{System Model}
\label{sec:2}
Consider a primary user transmitting OFDM signals with $\Nofdm$ subcarriers out of which only $K\ll\Nofdm$ are nonzero. A cognitive secondary user receives the transmitted signal of PU and wishes to detect empty subcarriers to start communicating over them. Furthermore suppose that because of relatively high cost and energy consumption of conventional Nyquist-based Analog-to-Digital Converters (ADC), the cognitive user samples the OFDM signal  at a sub-Nyquist rate to collect
compressive samples using an Analog-to-Information Converter (AIC) \cite{tropp10, kirolos06}. Matrix-wise, this can be described as
\beq
\vz[n] = \mA \vx[n],~n=1,\ldots,\Nb,
\label{eq1}
\eeq
where $\Nb$ is the number of collected OFDM symbols, $\vx[n]$ is the $n$-th noisy OFDM signal with unknown timing offset $d$, $\vz$ denotes the sub-Nyquist sampled signal, and $\mA$ is the $M \times N$ measurement matrix (notice that $K\le M<N$), where $N=\Nofdm+\Ncp$ and $\Ncp$ denotes the Cyclic Prefix (CP) length. The $n$-th OFDM symbol $\vx[n]$ can be written as
\beqa
\vx[n] =  \left[\begin{array}{c} \Ftilde_{d,2} \vs [n-1]  \\  \Ftilde_{d,1} \vs [n]\end{array}\right]+ \vw[n],~n=1,\ldots,\Nb,
\label{eq2}
\eeqa
where $\vs[n-1]$ and $\vs[n]$ are information vectors at time instants $n-1$ and $n$ which are both $K$-sparse with identical support set $\sS$. Moreover, $\Ftilde \triangleq [\mF_0,\mF]^H$ where $\mF$ is the $N_o \times N_o$ Discrete Fourier Transform (DFT) matrix defined as $[\mF]_{p,q} \triangleq \frac{1}{\sqrt{N_o}} e^{-j2\pi (p-1)(q-1)/N_o}$, $\mF_0$ is the submatrix of $\mF$ consisting of only the last $\Ncp$ columns of it, $\Ftilde_{d,1}$ is the submatrix of $\Ftilde$ consisting of its first $N-d$ rows and $\Ftilde_{d,2}$ is the submatrix of $\Ftilde$ consisting of its last $d$ rows. From (\ref{eq2}), the sub-Nyquist sampled signal at time instant $n$ can be written as 
\beqa
\vz[n]=\mA_{d,1} \Ftilde_{d,2} \vs [n-1]+\mA_{d,2} \Ftilde_{d,1}\vs [n]+\mA \vw[n],\nonr \\
~~~~~~~~~~~~~~~~~~~~~~~~~~~~~~~~~~~~n=1,\ldots,\Nb,
\label{eq3} 
\eeqa
where $\mA_{d,1}$ is the sub matrix of $\mA$ consisting of its first $d$ columns and $\mA_{d,2}$ is the sub matrix of $\mA$ consisting of its last $N-d$ columns. Let denote $\mBdy\triangleq\mA_{d,1} \Ftilde_{d,2}$ and $\mBdd\triangleq\mA_{d,2} \Ftilde_{d,1}$.
The set of $\Nb$ equations (\ref{eq3}) can be rewritten as the following comprehensive form
\beqa
\vz = \mB_d \vs + \vw,
\label{eq4} 
\eeqa
where 
\beqa
\mB_d \triangleq  \left[\begin{array}{ccccc} \mBdy& \mBdd & \mZero & \ldots & \mZero \\
\mZero & \mBdy & \mBdd & \ldots & \mZero \\
\vdots & \ddots & \ddots & \vdots & \mZero \\
\mZero &\mZero &\mZero & \ldots & \mBdd
\end{array}\right], 
\label{eq:equivMatFirst}
\eeqa
$\vz\triangleq [\vz[1]^T,\vz[2]^T,\ldots,\vz[\Nb]^T]^T$, and $\vs\triangleq [\vs[0]^T,\vs[1]^T,\ldots,\vs[\Nb]^T]^T$.

\section{Effects of timing offset}
\label{sec:delay}

Before proceeding with the proposed solution, we first investigate how the time delay in OFDM signal may affect the subcarrier identification process. 
We first remark that if there was no timing offset, i.e., $d=0$, then we could have estimated the joint support of vectors $\vs[n],~n=1,\ldots,\Nb$ by using the joint sparse recovery methods from multiple measurement vectors such as \cite{cotter05, chen06}. This is clear from (\ref{eq3}) as when $d=0$ the first term in the right-hand side of the equation disappears. But when $d>0$ we can no longer employ these algorithms as in this case each $\vs[n],~n=1,\ldots,\Nb-1$ appears in two measurement vectors $\vz[n-1]$ and $\vz[n]$ resulting these two measurements becoming correlated. 

Now consider the oracle-assisted case when we know the true set of active subcarrier indices $\sS$. Taking only the columns of $\mB_d$ and entries of $\vs$ corresponding to $\sS$ and discarding the rest,  (\ref{eq4}) can be rewritten as
\beq
\vz = \mBds \vs_\sS + \vw,
\label{eq42}
\eeq
where

\beqa
 \mBds \triangleq
 \begin{bmatrix}[c:c:c:c:c]
 \bovermat{$\mB_{d,\sS}^{(0)}$}{\mBdys}  & \bovermat{$\mB_{d,\sS}^{(1)}$}{\mBdds} & \bovermat{$\mB_{d,\sS}^{(2)}$}{~~~\mZero~~~}&
 {\cdots} & \bovermat{$\mB_{d,\sS}^{(\Nb)}$}{~~~\mZero~~~} \\
 \mZero & \mBdys & \mBdds & \ldots & \mZero  \\
\vdots & \ddots & \ddots & \vdots & \mZero \\
\mZero &\mZero &\mZero & \ldots & \mBdds \\
  \end{bmatrix}
\label{eq:equivMat}
 \eeqa
where $\mBdys \triangleq \mA_{d,1} \tilde{\mF}_{d,1,\sS}$, $\mBdds\triangleq \mA_{d,2} \tilde{\mF}_{d,1,\sS}$, and $\tilde{\mF}_{d,2,\sS}$ and $\tilde{\mF}_{d,1,\sS}$ are sub matrices of, respectively, $\tilde{\mF}_{d,2}$ and $\tilde{\mF}_{d,1}$ with columns corresponding to set $\sS$. Furthermore, $\vs_\sS \triangleq [\vs_\sS[0]^T,\vs_\sS[1]^T,\ldots,\vs_\sS[\Nb]^T]^T$ where $\vs_\sS[n],~n=0,\ldots,\Nb$ is the entries of vector $\vs[n]$ corresponding to set $\sS$ and $\mB_{d,\sS}^{(n)},~n=0,\ldots,\Nb,$ denotes the $(n+1)$-th block-column of $\mB_{d,\sS}$. From (\ref{eq:equivMat}) it is clear that vector $\vs_{\sS}$ can be recovered if $\mBds$ is full column rank. 
The following theorems state the necessary and sufficient conditions for recovering $\vs_\sS$.


{\thm A necessary condition for $\mBds$ having full column rank is that $K \le d \le N-K$.
{\proof
If $d < K$, then it is easy to see that $\rank(\mBdy) \le \rank(\mA_{d,1}) \le d < K$ and therefore $\rank(\mBdys)<K$ which means $\mBdys$ does not have full column rank. Since $\mBdys$ is the only nonzero block in the first block-column of $\mBds$ (see (\ref{eq:equivMat})), then matrix $\mBds$ does not have full column rank too. This concludes that in order for $\mBds$ to have full column rank we need to have $d \ge K$. 

Similarly, if $N-d < K$, then it is easy to see that $\rank(\mBdd)\le \rank(\mA_{d,2}) \le N-d<K$ and therefore $\rank(\mBdds)<K$ which mean that this time $\mBdds$ does not have full column rank. Since $\mBdds$ is the only nonzero block in the last block-column of $\mBds$, then matrix $\mBds$ does not have full column rank too. This concludes that in order for $\mBds$ to have full column rank we need to also have $d \le N-K$. \qed
}
 \label{thm2}}

{\thm Assume that $M\ge (\Nb+1)K/\Nb$ and entries of matrix $\mA$ are independent and continuous random variables. If $K \le d \le N-K$ then $\mBds$ will have full column rank with probability one. \label{thm1}}
{\proof The proof is deferred to Appendix. \qed}



The above theorems can be also explained with the following intuition: when $d<K$ the number of independent linear combinations of $K$ unknown nonzero elements of vector $\vs[0]$ appearing in the linear set of equation (\ref{eq42}), through linear mixture matrix $\mBdys$, is  $d<K$, which makes it impossible to estimate these elements. Similarly, when $N-d<K$, the number of independent linear combinations of $K$ unknown nonzero elements of vector $\vs[\Nb]$ appearing in the linear set of equation (\ref{eq42}), through linear mixture matrix $\mBdds$, is $N-d<K$, which again makes it impossible to estimate these elements. For the rest of the vectors $\{\vs[n]\}_{n=1}^{\Nb-1}$ this is not however the case as each of them appears twice in (\ref{eq42}): once through $\mBdys$ and once through $\mBdds$.

To solve the issue, when $d<K$, we cancel out the effect of unresolvable vector $\vs[0]$ by projecting $\vz[n],~n=1,\ldots,\Nb$ to the $(M-d)$-dimensional null space of $\mBdy$, denoted by $\mP_{\mBdy}^0$, to get 
\beqa
\acute{\vz}[n]&\triangleq& \nonr \mP_{\mBdy}^0 \vz[n] \\
&=&\mP_{\mBdy}^0 \mBdd \vs[n] + \mP_{\mBdy}^0 \mA \vw[n],~n=1,\ldots,\Nb.\nonr\\
\label{eq:s0case}
\eeqa
The projection in (\ref{eq:s0case}) causes us to lose the information lying on the $d$-dimensional subspace spanned by the columns of matrix $\mBdy$. This is not however a big loss as in this case $d$ is relatively a very small number. 


Similarly, when $N-d<K$, we cancel out the effect of unresolvable vector $\vs[\Nb]$ by projecting $\vz[n],~n=1,\ldots,\Nb$ to the $(M-N+d)$-dimensional null space of $\mBdd$, denoted by $\mP_{\mBdd}^0$, to get 
\beqa
\grave{\vz}[n]&\triangleq& \nonr \mP_{\mBdd}^0 \vz[n] \\
&=&\mP_{\mBdd}^0 \mBdy \vs[n-1] + \mP_{\mBdd}^0 \mA \vw[n],~n=1,\ldots,\Nb.\nonr\\
\label{eq:sNcase}
\eeqa
This time, the projection in (\ref{eq:sNcase}) causes us to lose the information lying on the $(N-d)$-dimensional subspace spanned by the columns of matrix $\mBdd$. Again, this is not a big loss as in this case  $N-d$ is relatively a very small number. 

\section{Active Subcarrier Identification}
\label{sec:3}

\subsection{Known timing offset}
\label{sec:known}
When the timing offset $d$ is known, active subcarriers can be identified by exploiting 
sparse reconstruction algorithms tailored to the model described in the previous two section. When $K \le d \le N-K$, we tailor the OMP algorithm \cite{tropp07} as follows to estimate the joint sparsity pattern of vectors  $\vs[n], n\in 0,\ldots,\Nb$: 

\begin{algorithm}
 \KwData{$\vz[n],~n=1,\ldots,\Nb$, $d$, $\mA$, $K$, set $\vz[0] = \mZero$}
 \KwResult{active subcarrier index set $\sS$}
 initialization: $k=0$, $\sS=\emptyset$\\
 \While{$k<K$}{
  $k \leftarrow k+1$,\\
  $\hat{j} = \arg \min\limits_{j\in\{1,\ldots,\Nofdm\}}  \sum\limits_{n=1}^{\Nb} \mbox{abs} \Big(\atilde_{d,j}^H \left[\begin{array}{c}\vz[n-1] \\ \vz[n]\end{array}\right]\Big)$,\\ 
  $\sS=\sS \cup \{\hat{j}\}$,\\
  Form $\mBds$ as in (\ref{eq:equivMat}),\\
  $\hat{\vs}_{\sS} = (\mBds^H \mBds)^{-1} \mBds^H \vz$,\\
  $\vz \leftarrow \vz-\mBds \hat{\vs}_{\sS}$,\\
  $\vz[n] \leftarrow \vz((n-1)M+1:nM),~n=1,\ldots,\Nb$,
   }
\caption{Proposed OMP-based joint sparse recovery algorithm for active subcarrier identification when $K \le d \le N-K$}
\end{algorithm}
The above algorithm re-confirms why the cases of $d<K$ and $d>N-K$ should be excluded, as $\mBds^H \mBds$ can be rank deficient in those cases.
Above $\atilde_{d,j}$ is the $j$-th column of matrix $\left[\begin{array}{c} \mBdd\\ \mBdy\end{array}\right]$.

On the other hand, for the two cases of $d<K$ and $d>N-K$, we can find the active subcarriers by applying the Simultaneous Orthogonal Matching Pursuit (SOMP) algorithm proposed in \cite{tropp06SOMP} to the set of $\Nb$ equations (\ref{eq:s0case}) and (\ref{eq:sNcase}), respectively. 



\subsection{Unknown timing offset}
\label{sec:unknown}
If the timing offset is unknown, we cannot solely resort to the algorithm introduced in the previous subsection to identify the active subcarriers. However, in this case active subcarriers and timing offset can be estimated by alternating between the following two steps:
\begin{itemize}
\item Step 1: Fix $d$ and estimate $\vs$ by solving the following optimization problem:
\beqa
\hat{\vs} = \arg \min_{\vs}  \| \vz - \mB_d \vs \|_2^2, \nonr \\
~~~\mbox{s.t.}~|\sS|=K
\eeqa
This step of the algorithm can be solved using the batch OMP algorithm introduced in Section \ref{sec:known}.
\item Step 2: Fix $\vs$ to the value estimated in Step 1 and estimate $d$ as
\beq
\hat{d}=\arg \min_{d}  \| \vz - \mB_d \vs \|_2^2,
\eeq
\end{itemize}
The above solution is in fact a joint dictionary learning and sparse approximation problem \cite{rubinstein10, ksvd, kreutz03}. In the first step, we fix the dictionary $\mB_d$ and estimate the joint sparse vectors $\vs[n], n\in 0,\ldots,\Nb$, and in the second step we fix the joint sparse vectors $\vs[n], n\in 0,\ldots,\Nb$ to the values estimated in the previous step and find the best estimate for dictionary $\mB_d$ from the dictionary set $\{\mB_0,\mB_1,\ldots,\mB_{N-1}\}$.


\section{Simulation Results and Analysis}
\label{sec:simulation}
In this section we study the performance of the proposed algorithm for joint estimation of delay and active subcarriers via some simulation studies.
For simplicity, we consider an OFDM system with IFFT size $32$ which means $\Nofdm=32$. The cyclic prefix length is chosen as $\Ncp=\Nofdm/4=8$ and the total OFDM symbol length is thus $N=\Nofdm+\Ncp=40$. Subcarrier symbols are assumed to be drawn from a 16-QAM constellation with unit energy. The elements of the measurement matrix $\mA$ are drawn from $\mathcal{N}(0,1)$ and its columns are normalized to have unit norm. We emphasize that these chosen values are just examples for carrying out numerical simulations, and are not as such related to the fundamentals of the proposed method in any way. 
Number of OFDM symbols taken for subcarrier identification is $\Nb=10$ or  $\Nb=20$. The number of random trials is 5000 to gather sufficient statistics for reliable evaluation of performance. The OFDM timing offset in each run is chosen uniformly randomly as an integer in the interval $[0,N-1]$. 

\begin{figure}[t!]
	\includegraphics[width=0.5\textwidth,height=0.35\textwidth]{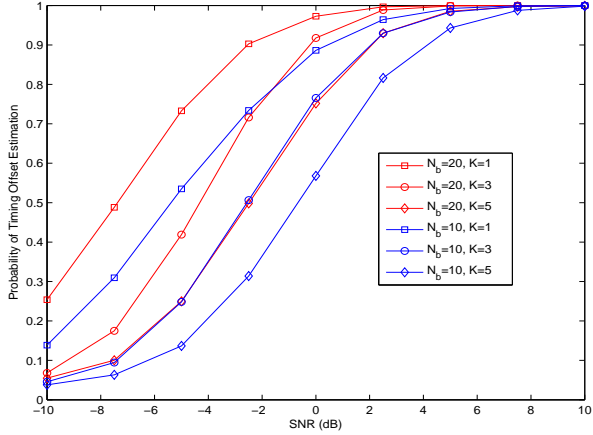} \vskip-0.3cm
	\caption{Probability of time delay estimation for an OFDM signal with number of active subcarriers $K \in \{1,3,5\}$, compression ratio $M/N=0.5$, and number of OFDM symbols $\Nb \in\{10,20\}$.} 
	\label{fig:delay1}
\vskip-0.2cm
\end{figure} 

\begin{figure}[t!]
	\includegraphics[width=0.5\textwidth,height=0.35\textwidth]{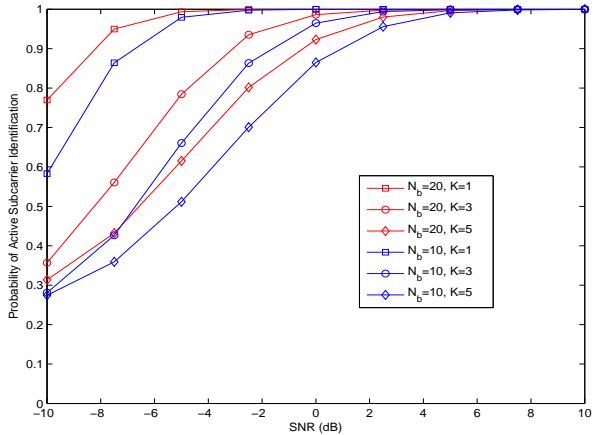} \vskip-0.3cm
	\caption{Probability of active subcarrier identification for an OFDM signal with number of active subcarriers $K \in \{1,3,5\}$, compression ratio $M/N=0.5$, and number of OFDM symbols $\Nb \in\{10,20\}$.} 
	\label{fig:active1}
\vskip-0.2cm
\end{figure}

In the first experiment we study the impact of the number of active subcarriers on both timing delay estimation and subcarrier identification. The compression ratio is set to $M/N=0.5$. We compute the results for three different choices of the number of active subcarriers $K \in \{1,3,5\}$ (chosen uniformly randomly within $\Nofdm$ total subcarriers) and two choices of the number of taken OFDM symbols $\Nb\in\{10,20\}$. Figure \ref{fig:delay1} shows the performance of the proposed algorithm for timing offset estimation of OFDM symbols. 
Moreover, Figure \ref{fig:active1} illustrates the performance of the proposed algorithm in terms of probability of correctly identifying active subcarriers.  The probability of active subcarrier identification is computed as the ratio of average of active subcarriers identified correctly over the total number of active subcarriers, where the average is taken over all random trials.
As we can see from Figures \ref{fig:delay1} and \ref{fig:active1} as the number of active subcarriers increases the OFDM signal becomes less sparse and therefore the performance of the sparse recovery algorithm which is a main part of the alternating method proposed in Section \ref{sec:unknown} decreases. Furthermore, as expected, the performance of the algorithm improves as the number of collected OFDM symbols increases. 

It is noteworthy to mention here that an error in estimation of the timing offset does not necessarily result in an error in identifying active subcarriers. This can be observed from comparing corresponding points in Figures \ref{fig:delay1} and \ref{fig:active1}. For example, at the signal-to-noise ratio of $-10$ dB for $K=1$ and $\Nb=20$, from Figure \ref{fig:delay1} it is seen that that the timing offset has been correctly estimated with a probability of about $0.25$, while Figure \ref{fig:active1} shows that the active subcarrier in this case has been correctly identified  with the probability of almost $0.77$. This is concisely because for small errors in estimation of $d$, there is still a big chance to detect the subcarriers correctly. This is, in turn, because for two timing offsets $d$ and $d^\prime=d+\epsilon$ where $\epsilon$ is a very small nonzero integer, there is a high correlation between a certain column of $\mB_d$ and the same column of $\mB_{d^\prime}$. A mathematical analysis of this observation will be given in an extended version of this paper.

\begin{figure}[t!]
	\includegraphics[width=0.5\textwidth,height=0.35\textwidth]{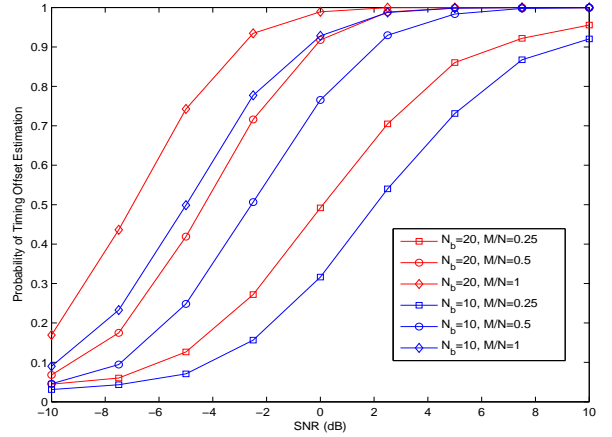} \vskip-0.3cm
	\caption{Probability of time delay estimation for an OFDM signal with compression ratio $M/N \in \{0.25,0.5,1\}$, number of active subcarriers $K=3$, and number of OFDM symbols $\Nb \in\{10,20\}$.} 
	\label{fig:delay2}
\vskip-0.2cm
\end{figure}

\begin{figure}[t!]
	\includegraphics[width=0.5\textwidth,height=0.35\textwidth]{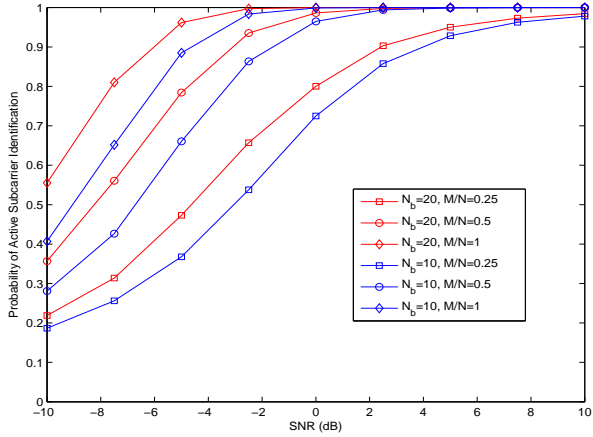} \vskip-0.3cm
	\caption{Probability of active subcarrier identification for an OFDM signal with compression ratio $M/N \in \{0.25,0.5,1\}$, number of active subcarriers $K=3$, and number of OFDM symbols $\Nb \in\{10,20\}$.} 
	\label{fig:active}
\vskip-0.2cm
\end{figure}

In the second experiment, the impact of the compression ratio on the performance of the proposed algorithm is studied. The number of active subcarriers is set to $K=3$ and we choose the compression ratio from set $M/N=\{0.25,0.5,1\}$. Figures \ref{fig:delay2} and \ref{fig:active} illustrate the performance of the proposed method in terms of probability of timing offset estimation and active subcarrier identification, respectively. As it can be observed the performance of the proposed method enhances as the compression ratio increases. This is also expected as the performance of the sparse recovery algorithm which is a main part of the alternating method proposed in Section \ref{sec:unknown} improves with the number of measurements $M$.

\section{Conclusion}
\label{sec:conclusion}
In this paper we proposed a technique for identification of sparse active subcarriers of an OFDM signal from sub-Nyquist samples in the presence of timing offset. We first studied the effect of timing offset and provided the sufficient and necessary conditions for signal recovery in the oracle-assisted case. Then, based on that, by tailoring the Orthogonal Matching Pursuit (OMP) algorithm, proposed a joint sparse recovery algorithm for identifying active subcarriers when the timing offset is known. Then we showed that for the case when the timing offset is unknown, the active subcarriers can be identified by alternating between a dictionary learning algorithm, which delivers an estimate for timing offset, and the joint compressive sensing algorithm proposed for known timing offset case. The method can be used for detecting spectral opportunities in cognitive radio systems. The performance of the proposed algorithm was studied via simulations.

\section*{Acknowledgement}
This work was supported by the Finnish Funding Agency for Technology and Innovation (Tekes) under the project "Enabling Methods for Dynamic Spectrum Access and Cognitive Radio (ENCOR)", and the Academy of Finland under the project $\#251138$ "Digitally-Enhanced RF for Cognitive Radio Devices".

\section*{Appendix: Proof of Theorem \ref{thm1}}

Before proving Theorem \ref{thm1}, we provide the following two lemmas which are used in the sequel to prove the theorem:

{\lem We have
\begin{enumerate}
\item $\spark(\tilde{\mF}_{d,2})=\min(N_0,d)+1$,
\item $\spark(\tilde{\mF}_{d,1})=\min(N_0,N-d)+1$.
\end{enumerate}
{\proof We only prove 1) here. The proof of 2) is similar. 

If $N_0<d$ then we have matrix $\tilde{\mF}_{d,2}=[\begin{array}{c} \acute{\mF} \\ \mF^H\end{array}]$ where $\acute{\mF}^H \triangleq \mF_{\{2 N_0-d+1:N_0\}}$. Since the $N_0$ columns of $\mF^H$  are linearly independent then so are the $N_0$ columns of $\tilde{\mF}_{d,2}$ and therefore 
\beq
\spark(\tilde{\mF}_{d,2})=N_0+1, ~\mathrm{when~} d >N_0.
\label{eq:appendix1}
\eeq 

On the other hand if $d \le N_0$ then $\tilde{\mF}_{d,1}=[\mF_{\{N_0-d+1:N_0\}}]^H$. From this it is easy to verify that
\beq
\tilde{\mF}_{d,1}=[\mF_{\{1:d\}}]^H \mLambda,
\eeq
where $\mLambda$ is a diagonal matrix whose $(i,i)$-th diagonal element is defined as $[\mLambda]_{i,i} \triangleq \omega_{N_0}^{-(N_0-d) i}$ where $\omega_{N_0} \triangleq e^{-j 2 \pi /N_0}$. Since $[\mF_{\{1:d\}}]^H$ is a full spark matrix \cite{fullspark}, therefore its product with the non-singular diagonal matrix $\mLambda$ is also full spark \cite{donoho03}; i.e. 
\beq
\spark(\tilde{\mF}_{d,1}) = d+1, ~\mathrm{when~} d \le N_0.
\label{eq:appendix2}
\eeq
The proof of Lemma is then concluded from (\ref{eq:appendix1}) and (\ref{eq:appendix2}).
\qed
}
\label{lem1}}

{\lem Consider the $Q \times P$ full column rank tall matrix $\mPhi$ and the $Q \times T$ random matrix $\mPsi$ whose entries are independent continuous random variables. Then with probability one we have
\beq
\mathrm{dim}\Big[\cC (\mPhi) \cap \cC(\mPsi)\Big]=P+\min(Q,T)-\min(Q,P+T).
\eeq
{\proof From \cite[Equation (2.17)]{marsaglia74} we have
\beqa
\mathrm{dim}\Big[\cC (\mPhi) \cap \cC(\mPsi)\Big] &=& \rank(\mPhi)+\rank(\mPsi)-\rank([\mPhi,\mPsi]) \nonr \\
&=&P+\min(Q,T)-\rank([\mPhi,\mPsi]).\nonr\\ && \label{eq:appp} 
\eeqa
If the entries of $\mPsi$ are independent continuous random variables, then with probability one its columns are linearly independent of themselves and also of the columns of $\mPhi$ as long as $P+T \le Q$ and therefore $\rank([\mPhi,\mPsi])=P+T$. When $P+T>Q$ on the other hand $\rank([\mPhi,\mPsi])=Q$. Putting these in (\ref{eq:appp}) the proof is concluded. \qed
}\label{lem2}}

Now we proceed to prove Theorem \ref{thm1}. 
First notice that matrix $\mBds$ is of size $M \Nb \times K(\Nb+1)$ which makes it a tall matrix since $M\ge (\Nb+1)K/\Nb$. Hence it is enough to prove that its columns are linearly independent. To this end, we first prove that if $K \le d \le N-K$ then both $\mBdys$ and $\mBdds$ in (\ref{eq:equivMat}) are full-column rank with probability one. 

We give the proof for $\mBdys$ here. The proof for $\mBdds$ is similar. From \cite[Chapter 4]{raorao} we know that
\beq
\rank(\mA_{d,1} \tilde{\mF}_{d,2,\sS}) = \rank(\mA_{d,1})-\mathrm{dim}[\cC(\mA_{d,1}^T) \cap \kK(\tilde{\mF}_{d,2,\sS}^T)] .
\label{eq:dametgarm1}
\eeq
Since entries of  $\mA$ are continuous random variables we have 
\beq
\rank(\mA_{d,1})=\min(M,d),
\label{eq:dametgarm2}
\eeq 
with probability one. Furthermore, from Lemma \ref{lem1}, we conclude that $\tilde{\mF}_{d,2,\sS}$ has full-column rank since $K<d$. Therefore we will have $\mathrm{dim}(\kK(\tilde{\mF}_{d,2,\sS}^T))=d-\mathrm{dim}(\cC(\tilde{\mF}_{d,2,\sS}))=d-\rank(\tilde{\mF}_{d,2,\sS})=d-K$. From these  and from Lemma \ref{lem2} we will have
\beqa
\mathrm{dim}[\cC(\mA_{d,1}^T) \cap \kK(\tilde{\mF}_{d,2,\sS}^T)]  = d-K+\min(M,d)- \nonr \\ \min(d,d+M-K) \nonr
\eeqa
Since $K<M$ (because of the sparsity condition) we will have $\min(d,d+M-K)=d$ and therefore
\beqa
\mathrm{dim}[\cC(\mA_{d,1}^T) \cap \kK(\tilde{\mF}_{d,2,\sS}^T)]  = \min(M,d)-K.
\label{eq:dametgarm3}
\eeqa
Inserting (\ref{eq:dametgarm2}) and (\ref{eq:dametgarm3}) in (\ref{eq:dametgarm1}) we will have 
\beq
\rank(\mBdys) \triangleq \rank(\mA_{d,1} \tilde{\mF}_{d,2,\sS})=K.
\label{eq:damamgarm1}
\eeq
In a similar fashion we can prove that 
\beq
\rank(\mBdds) \triangleq \rank(\mA_{d,2} \tilde{\mF}_{d,1,\sS})=K.
\label{eq:damamgarm2}
\eeq
Now, from \cite[Equation (8.3)]{marsaglia74} it is easy to see that $\rank([\mB_{d,\sS}^{(0)}~\mB_{d,\sS}^{(1)}])=2K$. Adding the block-columns of $\mB_{d,\sS}$ in (\ref{eq:equivMat}) one at a time and employing \cite[Equation (8.3)]{marsaglia74} yields
\beqa
\rank(\mB_{d,\sS}) &\triangleq& \rank([\mB_{d,\sS}^{(0)}~\mB_{d,\sS}^{(1)}~\ldots~\mB_{d,\sS}^{(\Nb)}]) \nonr \\
&=&(\Nb+1) K
\eeqa
which concludes the proof. \qed
\bibliographystyle{IEEEtran}

\end{document}